\documentclass[10pt, twocolumn]{article}

\usepackage{graphicx}

\usepackage{amsfonts,amsmath,pdfsync,color}
\usepackage{amsthm}
\usepackage{tikz}
\usepackage{adjustbox}
\usepackage{graphicx}
\usepackage{array}
\usepackage{appendix}
\usepackage{enumerate}
\usepackage{enumitem}
\usepackage{makecell}
\usepackage{wasysym}
\usepackage{multirow}
\usepackage{MnSymbol}
\usepackage{titling}
\usepackage[all]{xy}
\usepackage[intoc]{nomencl}
\usepackage{dsfont}
\usepackage[bookmarks=false]{hyperref}

\usepackage{authblk}

\title{Budget Allocation in Binary Opinion Dynamics}

\author[1]{Susana Iglesias Rey
\thanks{Email: \href{mailto:susana.iglesias_rey@nokia.com}{susana.iglesias\_rey@nokia.com}}}
\affil[1]{Nokia Bell Labs\\ Nokia Paris-Saclay\\ Route de Villejust\\ 91620 Nozay\\ France}
\author[2]{Patricio Reyes
\thanks{Email: \href{mailto:patricio.reyes@usc.es}{patricio.reyes@usc.es}}}
\affil[2]{
Technological Institute for Industrial Mathematics   (ITMATI)\\ University of Santiago de Compostela\\ Santiago de Compostela\\ Spain}
\author[1]{Alonso Silva
\thanks{Email: \href{mailto:alonso.silva@nokia-bell-labs.com}{alonso.silva@nokia-bell-labs.com} To whom correspondence should be addressed.}}
\date{}

\begin{document}
\maketitle
\begin{abstract}
In this article we study the allocation of a budget to promote an opinion in a group of agents. We assume that their opinion dynamics are based on the well-known voter model.
We are interested in finding the most efficient use of a budget over time in order to manipulate a social network. We address the problem using the theory of discounted Markov decision processes. Our contributions can be summarized as follows: (i) we introduce the discounted Markov decision process in our cases, (ii) we present the corresponding Bellman equations, and, (iii) we solve the Bellman equations via backward programming. This work is a step towards providing a solid formulation of the budget allocation in social networks.
\end{abstract}

\section{Introduction}
During the last decades a lot of research effort has been devoted to model the underlying process of opinion formation of agents that interact through a social network. In this respect, DeGroot's~\cite{DeGroot74} or Friedkin-Johnsen's \cite{FJ99} models are classic references. Another classic model is the voter model~\cite{Clifford73,Holley75} which considers that each agent holds a binary opinion, $0$ or $1$, and at each time step, each agent chooses one of its neighbors at random and adopts that opinion as its own. Other works, based on the voter model, incorporate stubborn agents~\cite{Yildiz13}, and biased agents~\cite{Arpan16}. Moreover, the last few years there has been an increasing literature about manipulation of opinion in social networks~\cite{Dhamal17,Forster13,Silva17}.

In this work, we are interested in finding the most efficient use over time of a budget in order to manipulate a social network. The idea is to promote an opinion by paying agents to supplant their true opinions. We model opinions as two values, 0 or~1, with 1 (0) representing supportive (non-supportive) opinion. 

We frame the problem of designing sequential payment strategies as a discounted Markov decision process (DMDP). DMDPs have been widely used to formulate many decision making problems in science and engineering (see, e.g.,~\cite{Altman02,Archak12,Boutilier16}). 
One of the main applications of DMDP models is the computation of optimal decisions (i.e., actions) over time to maximize the expected reward (analogously, minimize the expected cost).

First of all, we focus on a fully connected network where agents change their opinion following a voter model. We provide the correspondent Bellman equations to solve this problem and we show through an example how to solve the stated problem in practice. We provide a structural characterization of the associated value function and the optimal payment strategy. Then, we compute the optimal payment using dynamic backward programing.

\section{Model definition}
In order to find the optimal budget allocation on binary opinion dynamics, we make extensive use of the theory of DMDPs. First, we adopt the voter model of opinion formation over a social network. Then, we define the DMDP and the corresponding Bellman equations to obtain the optimal strategy of budget allocation. 

Consider an undirected social network $\mathcal{G}=(\mathcal{I}, \mathcal{E})$, where $\mathcal{I}$ stands for the set of agents, indexed from $1$ to~$n$, and $\mathcal{E}\subseteq \mathcal{I}\times\mathcal{I}$ is the set of edges. Each agent $i\in \mathcal{I}$ has a binary initial opinion.  Opinions take values $0$ or~$1$. If agent $i$ has opinion $0$ (analogously, $1$) we label it as \textit{non-supporter} (\textit{supporter}). For example, if agents are discussing about politics, $1$ could be supporting a particular party and $0$  not supporting it.

Moreover, we distinguish two cases, depending on whether the network is fully connected or not.

We start by studying the fully connected case. In each decision epoch,  agents update their opinions following a voter model sensitive to external payments. Let $\beta\in [0, 1)$ be the discount factor, which represents the loss of reward in the future with respect to the current reward. A discounted Markov decision process (DMDP) is a 5-tuple
$(S, A_S, \mathcal{P}, R, \beta)$, where $S$ is a finite set of states, $A_{S}$ is a finite set of actions, $\mathcal{P}$ is the set of transition probabilities and $R$ is the set of rewards. 

Therefore, the model is defined as follows:
\begin{enumerate}[noitemsep,topsep=0pt]
\item Decision epochs: the set of decision epochs is defined as $t = 1, 2,\ldots,T$. We consider a finite discrete-time system. At each decision epoch $t$ we observe the state $s$ and choose an action $a$.
\item States: the state space $S$ of the DMDP consists on the possible number of supporters, $s \in \mathbb{N}$, \break $s=|\{i\in \mathcal{I} \mid i \text{ is a supporter}\}|$.
\item Actions: the action space $A_s$ is the set of actions available in state $s$ (without loss of generality, actions are state independent). We consider that the actions $a \in \mathbb{N}$ are the possible number of payments, $a=|\{i\in\mathcal{I}\mid i \text{ receives a payment}\}|$, where the non-supporter agents have a cost $c_{NS}$ for changing their opinion from non-supporter to supporter, and the supporter agents, a cost $c_{S}$ to hold their supporter opinion. We assume that the cost of changing their opinion is higher than the cost of holding it, i.e., $c_{NS}>c_{S}$. We consider a finite budget $b$. Notice that, because the actions are constrained by the budget, they are stationary.
\item Transition probabilities: if the DMDP in decision epoch $t$ is at state $s_1$, the probability that it transitions to state $s_2$ taking action $a$ is denoted $p_t(s_1, s_2, a)$. Due to the natural independence of agents transitions, we compute those probabilities as the product of the transition probabilities of the agents. The evolution of one agent $i\in \mathcal{I}$ will be described by the voter model. Starting from any arbitrary initial labels, supporter (S) or non-supporter (NS), we consider two labeling functions $f_{S}(i)$ and $f_{NS}(i)$, where $f_S(i)=1$ ($f_{NS}(i)=1$) means that agent $i$ is a supporter (non-supporter). At each decision epoch $t$, each node selects uniformly at random one of its neighbors opinion. For each node $j \in\mathcal{I}$, the set of its neighbors is defined as $N(j) = \{k \in  \mathcal{I} \mid  \{j, k\} \in \mathcal{E}\}$. Therefore, we define for one $i\in \mathcal{I}$ with zero-payment in decision epoch $t$ the labeling functions,
\[f_S^{t+1}(i)=
\begin{cases}
1 \text{ with prob. } \frac{|\{j\in N(i) \mid f_S^{t}(j)=1 \}|}{|N(i)|},\\
0 \text{ with prob. } \frac{|\{j\in N(i) \mid f_S^{t}(j)=0 \}|}{|N(i)|},
\end{cases}\]
\[f_{NS}^{t+1}(i)=1 - f_S^{t+1}(i)
\]
Analogously, for one agent $k\in \mathcal{I}$ that receives a payment in decision epoch $t$ we define:
\[f_S^{t+1}(k)=
\begin{cases}
1 \text{ with prob. } 1,\\
0 \text{ with prob. } 0,
\end{cases}\]
\[f_{NS}^{t+1}(k)=1-f_S^{t+1}(k)
\]

As we said, we assume that the graph is fully connected, therefore each agent can communicate with every other agent. We denote the set of non-supporter agents that receive a payment as $\mathcal{L}=\{i\in \mathcal{I}\mid\text{ non-supporter}\ \&\ \text{ receives a payment}\}$ and its cardinality as  $\ell=|\mathcal{L}|$. Respectively, the set of supporter agents that receive a payment as $\mathcal{K}=\{i\in \mathcal{I}\mid \text{ supporter}\ \&\ \text{ receives a payment}\}$ and its cardinality as $k=|\mathcal{K}|$. Notice that~$a=\ell+k$.

Therefore the transition probabilities $p_t(s_1,s_2,a)$ can be computed as:
\begin{itemize}[noitemsep,topsep=0pt,label=\raisebox{0.2ex}{\tiny$\bullet$}]
\item $\text{If }  a>s_2 \text{ then } p_t(s_1, s_2, a)=0.$
\item $\text{If }  a<s_2 \text{ then }  $
\end{itemize}
\end{enumerate}
{\small\begin{align*}
&p_t(s_1, s_2, a) =\\
&\sum_{i=\max\{0, s_2-(n-s_1)-k\}}^{\min\{s_1-k, s_2-\ell\}} \binom{s_1-k}{i}\left(\frac{s_1}{n}\right)^i
\left(\frac{n-s_1}{n}\right)^{s_1-k-i}\\
&\cdot \binom{n-s_1-\ell}{s_2-i-(\ell+k)}\cdot \left(\frac{n-s_1}{n}\right)^{n-s_1-s_2+i+k}\left(\frac{s_1}{n}\right)^{s_2-i-(\ell+k)}
\end{align*}
}
\begin{enumerate}\addtocounter{enumi}{4}
\item Reward: the instant reward in time $t$ and state $s$ is defined as $r_t(s)=\sum_{i\in \mathcal{I}}g\cdot s $, where $g$ denotes the reward provided by one agent. 
\end{enumerate}

Let $\upsilon_{\beta}$ be the value function of the above DMDP, i.e., it is the supreme, over all possible budget allocation strategies, of the expectation of the discounted reward starting from an initial budget $b$. Under these assumptions, the Bellman equations for all $s\in S$ and initial budget $b$ are:\begin{align*}
\upsilon_{\beta}(s,b)& = \max_{\substack{a=\ell+k\in A_{S},\\ (\ell\cdot c_{NS} + k \cdot c_{S}) \leq b}} \mathbb{E}\sum_{t=0}^{\infty}\beta^t r_t(s) \\
 = &\max_{\substack{a=\ell+k\in A_{S},\\ (\ell\cdot c_{NS} + k \cdot c_{S}) \leq b}} \Big\{r_0(s) + \sum_{s'\in S}\beta p_1(s, s', a) \\
&\cdot \upsilon_\beta(s',b-\ell\cdot c_{NS} - k \cdot c_{S}+  r_0(s))\Big\},
\end{align*}
where the budget evolves as 

\mbox{$b(t+1)=b(t) + r_t(s)- \ell\cdot c_{NS} - k \cdot c_{S}.$}\\

Next, we present the second case where in each decision epoch  agents update their opinions following a voter model in a  network (not necessarily fully connected) that can be affected by external payments. As before, we design this problem as a DMDP where the set of actions are the possible external payments.

Therefore, let $\beta\in [0, 1)$ be the discount factor, we consider, by a slight abuse of notation, the 5-tuple $(S, A_S, \mathcal{P}, R, \beta)$ as before.
Concretely, the elements changed from the previous model:
\begin{enumerate}[noitemsep,topsep=0pt]
\item Decision epochs:  the set of decision epochs is defined as $t = 1, 2,\ldots,T$.
\item States: the state space $S$ of the DMDP, consists on all possible combinations of agents' labels, non-supporter ($0$) or supporter ($1$), i.e., $s \in \{0, 1\}^n$.
\item Actions: the action space $A_s$ is the set of actions available in state $s$ (without loss of generality, actions are state independent). An action means whether or not we give a payment to each of the agents, $a \in \{0, 1\}^n$, where a $0$ (respectively $1$) in position $i$ means we give no payment (payment) to agent $i\in\mathcal{I}$. We also define a vector of costs $c\in\mathbb{R}_+^n$ whose element $c_i>0$ is the cost of changing by one unit the opinion of agent $i$, in case agent $i$ is non-supporter, or the cost of holding the opinion of agent $i$, in case agent $i$ is a supporter. 
\item Transition probabilities: as before, if the DMDP in decision epoch $t$ is at state $s_1$, the probability that it transitions to state $s_2$ taking action $a$ is expressed as $p_t(s_1, s_2, a)$ and can be computed as:
\begin{itemize}[noitemsep,topsep=0pt,label=\raisebox{0.2ex}{\tiny$\bullet$}]
\item $\text{If }  |a|>|s_2| \text{ then } p_t(s_1, s_2, a)=0.$
\item $\text{If }  |a|<|s_2| \text{ then } p_t(s_1, s_2, a) \text{ is equal to }${\small\begin{alignat*}{2}
&\prod_{i=1}^{n} \mathbb{I}\{&&s_i=s_i'\}\Big[\mathbb{I}\{s_i=\text{supp}\}\frac{|\{j\in N(i) / f_{S}^{t}(j)=1 \}|}{|N(i)|}\\
& &&+\mathbb{I}\{s_i=\text{non-supp}\}\frac{|\{j\in N(i) / f_{NS}^{t}(j)=1 \}|}{|N(i)|}\Big] \\
&  \hspace{2mm}+ \mathbb{I}\{&&s_i\neq s_i'\}\Big[\mathbb{I}\{s_i=\text{supp}\frac{|\{j\in N(i) / f_{S}^{t}(j)=0 \}|}{|N(i)|}\\
& &&+ \mathbb{I}\{s_i=\text{non-supp}\}\frac{|\{j\in N(i) / f_{NS}^{t}(j)=0 \}|}{|N(i)|}\Big].
\end{alignat*}}
\end{itemize}
\item Reward: the instant reward in time $t$ and state $s$ is defined as $r_t(s)=g^T \cdot s$, where $g$ is the vector of rewards whose element $g_i$ is the reward that  agent $i\in\mathcal{I}$ provides.
\end{enumerate}

Let $\upsilon_{\beta}$ be the value function of the above DMDP, i.e., it is the supreme, over all possible budget allocations strategies, of the expectation of the discounted reward starting from an initial budget $b$. Under these assumptions the Bellman equations for all $s\in S$ and initial budget $b$ are:
{\small\begin{align*}
&\upsilon_{\beta}(s,b) = \max_{\substack{a\in A_{S},\\  c^T \cdot a \leq b}}\mathbb{E}\sum_{t=0}^{\infty}\beta^t r_t(s) \\
& = \max_{\substack{a\in A_{S},\\  c^T \cdot a \leq b}} \Big\{r_0(s) + \sum_{s'\in S}\beta p_1(s, s', a) \upsilon_\beta(s',b- c^T \cdot a+  r_0(s))\Big\},
\end{align*}}where the budget evolves as $b(t+1)=b(t)- c^T\cdot a + R(s_t)$ and $c^T$ denotes the transpose of vector $c$. 

\section{Simulation results}

We suppose that after a time $T$ we will not obtain rewards for the supporter agents, so we are interested in the distribution of the DMDP in the time interval $[0, T]$. We consider an undirected, fully connected social network $\mathcal{G}=(\mathcal{I}, \mathcal{E})$ with $n=7$ agents that form opinions with a voter model. In time $t=0$, we assign at random an initial label to each agent. Solving the Bellman (fixed point) equations, given the model and the set of states and feasible actions, gives us the best strategy for each state to follow in the time interval. We take $T=6$, $\beta=0.8$, $c_{NS}=10$, $c_{S}=5$, $B=30$ and $g=8$.

Some conclusions can be drawn from the simulations. The optimal payment strategy is to invest all our budget paying to the higher number of agents at time $t=0$. Obviously, if all the network is non-supporter (respectively supporter), the budget will be allocated to change opinions (to hold opinions). However, the distribution of the budget differs for the rest of possible initial states as we show on Table \ref{Table}.

\begin{table}[!ht]
\centering
\begin{tabular}{|c|cc|}
\hline
\multirow{2}{*}{Initial state} & \multicolumn{2}{c|}{Budget Allocation} \\
& Payments to NS & Payments to S \\
\hline
$s=1$ & 2 & 1\\
$s=2$ & 2 & 2\\
$s=3$ & 1 & 3\\
$s=4$ & 1 & 4\\
$s=5$ & 0 & 5\\
$s=6$ & 0 & 6\\
\hline
\end{tabular}
\caption{Budget allocation over the states.\label{Table}}
\end{table}

\begin{figure}[!ht]
\centering
\includegraphics[width=7cm]{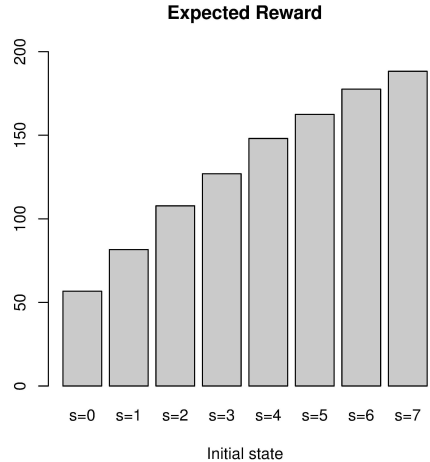}
\caption{Expected reward at time $T$. \label{Figure1}}
\end{figure}

Given the budget allocation at time $t=0$, we show in Figure \ref{Figure1} the expected reward obtained at time $T$ for each state $s\in S$. 

\section{Conclusions}

In this work, we have introduced the problem of budget allocation over time to manipulate a social network. We have developed a formulation of the discounted Markov decision process as well as the corresponding Bellman equations for the voter model of opinion formation. Using backward programming, we have obtained the optimal payment strategy for a small example of agents interacting trough a fully connected network. Many questions still remain to be answered. Future work would be devoted to improve the performance of our simulations in order to obtain the optimal strategy for larger networks and different topologies. Moreover, we intend to construct the DMDP model and the Bellman equations for different models of opinion dynamics. This will lead to the mathematical characterization of the optimal policy for different network structures and opinion formation models. It will lead also to the characterization of the most \textit{important agents} (the agents with highest benefit-cost ratio) which should be related with its centrality as shown in previous results \cite{Silva17,Dhamal17}.  \\

\section*{Acknowledgements}
The work of A.~Silva and S.~Iglesias Rey was partially carried out at LINCS (\url{www.lincs.fr}).

\bibliographystyle{hieeetr}
\bibliography{bibliography}

\end{document}